\documentclass[iop,apj,onecolumn]{emulateapj}

\usepackage{amsmath}

\newcommand{\eq}{Equation~}
\newcommand{\eqs}{Equations~}
\newcommand{\Fig}{Figure~}

\newcommand{\fig}{Figure~}
\newcommand{\figs}{Figures~}

\newcommand{\ie}{i.e.,}
\newcommand{\eg}{{e.g.,}}

\newcommand{\dd}{\mathrm{d}}
\newcommand{\nablab}{\boldsymbol{\nabla}}
\newcommand{\uT}{u_{x\mathrm{T}}}
\newcommand{\vT}{u_{y\mathrm{T}}}
\newcommand{\bu}{\mathbf{u}}
\newcommand{\bB}{\mathbf{B}}
\renewcommand{\Pr}{\mathrm{Pr}}

\begin{document}

\title{\uppercase{A self-consistent model of the solar tachocline}}
\author{\sc T.~S.~Wood$^{1}$ and N.~H.~Brummell$^{2}$}
\affil{$^1$School of Mathematics, Statistics and Physics, Newcastle University, United Kingdom}
\affil{$^2$Department of Applied Mathematics and Statistics, Baskin School of Engineering, University of California Santa Cruz}
\email{toby.wood@ncl.ac.uk}

\begin{abstract}
  We present a local but fully nonlinear model of the solar tachocline, using three-dimensional direct numerical simulations.
  The tachocline forms naturally as a statistically steady balance between Coriolis, pressure,
  buoyancy and Lorentz forces beneath a turbulent convection zone.
  Uniform rotation is maintained in the radiation zone by a primordial magnetic field,
  which is confined by meridional flows in the tachocline and convection zone.
  Such balanced dynamics has previously been found in idealised laminar models,
  but never in fully self-consistent numerical simulations.
\end{abstract}


\section{\uppercase{Introduction}}

\defcitealias{WoodBrummell12}{WB12}%

\subsection{The importance of the tachocline for solar models}
Though it's existence has been known for more than 30 years, a completely self-consistent model of
the solar tachocline is still lacking.
This is despite the crucial importance of the tachocline in models of the solar magnetic cycle
\citep[\eg][]{SpiegelWeiss80,WangSheeley91,Parker93}.
Traditionally, numerical simulations of the solar dynamo have modeled the convection zone in isolation
\citep[\eg][]{GilmanMiller81,Glatzmaier84,Miesch-etal08},
in part because attempts to include the radiation zone in these models often lead to un-solar-like results
\citep[\eg][]{Miesch-etal00}.
The absence of a self-consistent and \emph{realistically thin} tachocline in these models
may explain why
they
have difficulty producing solar-like magnetic cycles
\citep[\eg][]{Browning-etal06}.
Although some models do generate dipolar magnetic fields with regular reversals
\citep[\eg][]{Kapyla-etal12,PassosCharbonneau14,Strugarek-etal17},
they lack the smaller-scale features that characterise the solar dynamo,
such as the coherent magnetic flux tubes that form sunspots and magnetic prominences,
which are generally believed to originate in the tachocline \citep{Parker55-buoyancy}.
The tachocline may also influence the dynamics of the convection zone
in other ways.  For example, the presence of a latitudinal entropy gradient in the tachocline may help
to explain not only the differential rotation of the tachocline itself \citep{GoughMcIntyre98},
but also that of the convection zone \citep{Rempel05,Miesch-etal06,Balbus-etal12}.

The most remarkable and puzzling feature of the tachocline is that its thickness ---
inferred from
helioseismology ---
is less than $4$\% of the solar radius \citep{BasuAntia03}.
As first recognized by \citet{SpiegelZahn92}, this implies that angular momentum transport
in the tachocline must be predominantly horizontal, and also frictional (\ie\ down-gradient in angular velocity).
The source of this transport could be either horizontal turbulence \citep{Zahn92,SpiegelZahn92}
or the Maxwell stress from a primordial magnetic field \citep{RudigerKitchatinov97,GoughMcIntyre98}.
The main difficulty with the turbulence explanation is that transport by horizontal turbulence is not generally
frictional \citep[\eg][]{McIntyre94,GoughMcIntyre98,Tobias-etal07}.
Moreover, even if we suppose that horizontal turbulence can explain the thinness of the tachocline,
we must then invoke an additional mechanism to explain the uniform rotation of the deep radiation zone,
such as angular momentum extraction by internal waves \citep[\eg][]{KumarQuataert97,Zahn-etal97}.

The advantage of the magnetic explanation is that it explains both the thinness of the tachocline
and the uniform rotation of the radiation zone,
\emph{provided that the field remains confined below the convection zone},
meaning that the field lines do not extend across the tachocline into the convection zone.
An unconfined field, on the other hand, would imprint the differential rotation of the convection zone onto
the radiation zone through the Alfv\'enic elasticity of the magnetic field lines
\citep{Ferraro37,MacGregorCharbonneau99}.
The problem, then, is to explain why the field should remain confined to the radiation zone.

\begin{figure}[h!]
  \centering%
  \includegraphics{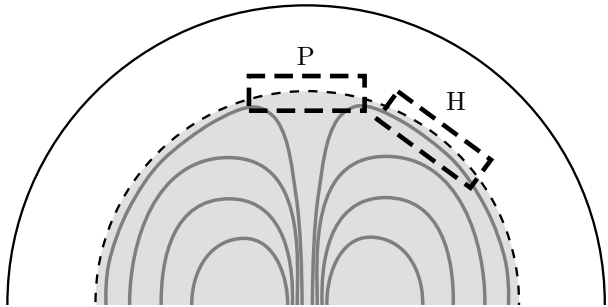}%
  \caption{%
    A global-scale primordial magnetic field (shown in gray), confined within the radiation zone (lightly shaded),
    can enforce uniform rotation in that region.
    The dashed boxes indicate the high-latitude (H) and polar (P) regions that we model numerically.}
  \label{fig:global}%
\end{figure}

\subsection{The \citeauthor{GoughMcIntyre98} model}
\label{sec:GM98}
The first study to directly address the magnetic confinement problem was that of \citet{GoughMcIntyre98}.
Previous studies \citep{MestelWeiss87,RudigerKitchatinov97} had taken the confinement of the field for granted,
assuming that the field would be expelled from the convection zone by convective turbulence
\citep{Zeldovich57,Weiss66}.
However, although the ability of turbulence to confine a horizontal magnetic field has been convincingly demonstrated
in numerical simulations
\citep[\eg][]{Nordlund-etal92,Tao-etal98,Tobias-etal98}, it is less clear that turbulence can confine a \emph{vertical}
magnetic field.
If the Sun's primordial field has an axial dipolar geometry, as suggested in \fig\ref{fig:global},
then the vertical component of the field will be strongest close to the poles.
On this basis, \citeauthor{GoughMcIntyre98} argued that the polar magnetic field can only be confined
by meridional flows that downwell in the tachocline and hold the field in an essentially laminar advection--diffusion balance.
Such downwelling meridional flows are, in fact, expected in the high-latitude tachocline,
as a result of ``gyroscopic pumping'' by the retrograde rotation of the overlying convection zone
\citep{SpiegelZahn92,McIntyre00,WoodMcIntyre11,WoodBrummell12,Miesch-etal12}.
The characteristic velocity of this downwelling,
$U$, say,
can be estimated from the observed differential rotation
of the tachocline, if we assume that the tachocline is in thermal-wind balance and local thermal equilibrium
\citep[p.~194]{GoughMcIntyre98,McIntyre07}.  In this way, \citeauthor{GoughMcIntyre98} estimated a value of
$U \simeq 10^{-5}$cm\,s$^{-1}$, sufficient to confine the magnetic field across an extremely thin ``tachopause''
at the bottom of the tachocline.  The thickness of the tachopause
in their model is $\eta/U \simeq 4\times10^7$cm,
where $\eta \simeq 400\,$cm$^2$\,s$^{-1}$ is the magnetic diffusivity of the tachocline.
This is only a tiny fraction of the tachocline thickness, $0.04R_\odot \simeq 3\times10^9$cm.

The \citeauthor{GoughMcIntyre98} tachocline model is characterized by
advection--diffusion balances in the induction and heat equations,
a balance between Coriolis, pressure gradient, and buoyancy forces
in the meridional directions,
and a balance between Coriolis and Lorentz forces in the azimuthal direction.
Several attempts have been made to reproduce this model in self-consistent, fully non-linear
direct numerical simulations,
but such balanced dynamics have never been obtained.
Instead, the transport of angular momentum in such simulations is generally dominated by viscosity,
and the magnetic field is found to diffuse out of the radiation zone and become unconfined
\citep{BrunZahn06,Rogers11,Strugarek-etal11}.
Under these conditions a significant shear between the convection and radiation zones can only be maintained
on a relatively short timescale, and eventually the differential rotation of the convection zone must
spread into the radiation zone
\citep{Brun-etal11}.
Of course, since numerical simulations cannot be performed at the true parameter values of the solar interior,
these results may simply reflect the fact that the simulations have not been performed in the correct parameter regime.
As discussed by \citet{Wood-etal11},
the lack of field confinement in these simulations is probably explained, at least in part,
by the predominance of viscosity in the dynamics,
which inhibits the burrowing of meridional flows into the radiation zone.
By contrast, the dynamics described by \citet{GoughMcIntyre98}
are essentially inviscid.

To determine whether or not the \citeauthor{GoughMcIntyre98} model is truly applicable to the solar tachocline,
we must therefore determine the conditions under which viscosity
does not play a significant role in the dynamics.
Despite a considerable literature on the subject \citep[\eg][]{Vogt25,Eddington25,Sweet50,Mestel53,Howard-etal67,Sakurai70,Spiegel72,Clark73,Clark75,Osaki82,Haynes-etal91,SpiegelZahn92,Elliott97,McIntyre02,GaraudBrummell08},
until recently no direct numerical simulation had ever
achieved the dynamical regime believed to operate in the tachocline,
in which the transport of angular momentum by meridional flows
dominates the transport by viscosity.
This dynamical regime is typical
in astrophysical objects, but is very difficult to achieve in numerical simulations
\citep[see][and references therein]{WoodBrummell12}.
The most crucial condition is that the turnover time for the meridional circulation must be shorter than
the viscous diffusion time across the same region.  For a strongly stably stratified fluid like the tachocline,
this condition can be expressed roughly as $\sigma < 1$, where
\begin{equation}
  \sigma = \frac{N}{2\Omega}\Pr^{1/2}.
\end{equation}
Here $N$ is the buoyancy frequency, $\Omega$ is the mean rotation rate, and $\Pr$ is the Prandtl number
--- the ratio of viscous and thermal diffusivities.
The tachocline has $N/2\Omega \simeq 150$ and $\Pr \simeq 2\times10^{-6}$,
and so $\sigma \simeq 0.2$ \citep{GaraudAA09}.
Most numerical models of the solar interior use realistic values for $N$ and $\Omega$ but,
owing to computational limitations, use values of $\Pr$ that are much closer to unity.
As a result, these simulations have $\sigma \gg 1$, leading to dynamics that are dominated by viscosity.
In order to achieve the correct ``low-sigma'' regime in numerical simulations, it is necessary to use
non-solar values of either $N$ or $\Omega$.
To our knowledge, the first fully nonlinear simulations ever performed in the correct ``low-sigma'' regime were those
of \citet{WoodMcIntyre11}, who used a Boussinesq, cylindrical code to model the polar tachocline.
However, their model was laminar and axisymmetric, and was only intended to model the bottom of the tachocline
--- the tachopause of \citeauthor{GoughMcIntyre98}.
Subsequently, \citet{WoodBrummell12},
hereafter referred to as \citetalias{WoodBrummell12},
used a fully compressible, local Cartesian code
to study the driving of meridional flows in the radiation zone by the differential rotation of the convention zone.
For the first time, these simulations demonstrated that the ``radiative spreading'' of differential rotation by
burrowing meridional flows described by \citet{SpiegelZahn92} can operate in a self-consistent, fully nonlinear
model including the generation of internal waves by convective overshoot.

Although the simulations of \citetalias{WoodBrummell12}
did not include a magnetic field, the meridional flows obtained closely resembled
those anticipated by the \citeauthor{GoughMcIntyre98} model, suggesting that magnetic field confinement might be possible.
The purpose of the present paper is to test this hypothesis.
We use the same local Cartesian model, but add a magnetic field within the radiation zone, in order see whether
the field can be confined, and whether a thin tachocline is then obtained.
We emphasize that the magnetic field considered here is of primordial, not dynamo, origin.
Obtaining a self-consistent solar dynamo is beyond the scope of the present study and, in any case,
it seems likely that a self-consistent tachocline model is a prerequisite for obtaining a realistic solar dynamo
\citep[\eg][]{Browning-etal06}.

The timescale for ohmic decay of a primordial magnetic field in the radiation zone is of the order of a billion years.
This is somewhat shorter than the age of the Sun, implying that at the present age the remaining field must resemble
a global-scale dipole, but much longer than the dynamical timescale of the tachocline, and so the field
in the radiation zone can be regarded as steady on the timescales of interest here.
In our local model we will therefore maintain a magnetic field by adopting suitable boundary conditions,
and we will choose the geometry of the field to represent different latitudes within the tachocline.

\subsection{Selection of parameter values}

It is crucial to choose the physical parameters of the problem appropriately,
and so we take guidance from the results of previous studies.
The hydrodynamic (\ie\ non-magnetic) parameters are chosen to match those of \citetalias{WoodBrummell12}
in a case where they found the correct behavior for meridional flows.
In choosing the magnetic parameters, i.e., the field strength and magnetic diffusivity,
we are guided by the results of previous, idealized models
\citep{WoodMcIntyre11,Wood-etal11,AA-etal13}.
Each of these models predicted that, for a certain range of magnetic field strengths,
a primordial magnetic field can indeed be confined beneath the convection zone,
resulting in a thin tachocline.  However, each of these models considered only an axisymmetric steady-state
balance, and so the effects of turbulence and waves were either parameterized or else neglected completely.
Here, we will solve the full set of three-dimensional, nonlinear, compressible MHD equations self-consistently.

\section{\uppercase{The Numerical Model}}
\label{sec:model}

The computational model used here is an extension of that used in \citetalias{WoodBrummell12},
which is based on the compressible $f$-plane code of \citet{Brummell-etal02}.
The code solves
the ideal gas equations in a Cartesian box using a rotating frame of reference.
The computational domain comprises two layers: an upper, convectively
unstable (and therefore turbulent) layer, and a lower, stably stratified layer.
The transition from stable to unstable stratification is produced by a change in the
thermal conductivity, $k$, which is prescribed as a function of depth, $z$.
In all simulations this transition is located at the mid-height of the domain.
For simplicity, as in \citetalias{WoodBrummell12}, we take the rotation axis to be vertical
($\boldsymbol{\Omega} = -\Omega\mathbf{e}_z$).
The main difference between the model used here and that of \citetalias{WoodBrummell12}
is the inclusion of magnetic fields, which influence the dynamics through the Lorentz force
and, to a much lesser extent, ohmic heating.

In order to describe the effect of the global-scale field in our local-scale model,
we consider two different configurations for the magnetic field,
which approximate the topology of the Sun's primordial field at different latitudes.
We refer to these two field configurations as ``horizontal'' (H) and ``polar'' (P),
as indicated in \fig\ref{fig:global}.
The horizontal field configuration represents conditions within the tachocline in high latitudes,
but away from the pole, where the field in the radiation zone is approximately horizontal,
whereas the polar field configuration represents conditions in a neighborhood of the pole,
where the field is more vertical.
We adopt Cartesian
coordinates in which
$z$ corresponds to depth;
in the horizontal field simulations,
$x$ and $y$ correspond to
azimuth and colatitude, respectively.
In all simulations, the computational domain
is periodic in both $x$ and $y$.
In order to implement the polar field configuration with periodic boundary conditions,
the simulations actually include four magnetic poles,
using a Peirce quincuncial projection,
as illustrated in \fig\ref{fig:polar-initial}.
The results we present in Section~\ref{sec:polar} are averaged over the four poles.
The boundary conditions employed for the magnetic field in the horizontal and polar
configurations are described in detail in Section~\ref{sec:BCs} below.

The horizontal field simulations have a box size of $(2L)^3$,
where $L$ is the thickness of the convective layer,
and a numerical resolution corresponding to $100\times100\times200$ grid points.
The polar field simulations have a box size of $8L\times8L\times2L$,
and a numerical resolution of $200^3$ grid points.
As in \citetalias{WoodBrummell12}, the requirement of small Prandtl number, $\Pr$,
places a severe constraint on the maximum permitted computational time step.
Moreover, the simulations typically must be continued for at least a domain-scale magnetic diffusion
time in order for the magnetic field to reach a statistically steady state.
Each of the simulations presented in this paper was run for more than 5 million numerical time steps.
These computational constraints preclude a higher spatial resolution.

\subsection{Boundary and Initial Conditions}
\label{sec:BCs}
We impose constant temperature $T_0$ at the top of the
domain, $z=0$, and a constant upward heat flux $H = -k\,\dd T/\dd z$ at the bottom,
$z=2L$.
The fluid is initially at rest and in hydrostatic balance
with uniform vertical heat flux throughout,
and has pressure $p_0$ and density $\rho_0$ at the upper boundary, $z=0$.
The top and bottom boundaries of the domain
are modeled as impenetrable
and stress-free.
This ensures that the total mass of fluid
in the domain does not change during the simulation,
and that no viscous torque is exerted at the boundaries.

The remaining boundary conditions are chosen to allow the system to achieve a
statistically steady state with a finite mean magnetic field.
For the horizontal field simulations, we impose ``perfectly conducting'' boundary conditions
at the top and bottom boundaries,
meaning that $\nablab\times\bB$ is perpendicular to the boundaries.
Although these boundary conditions are somewhat artificial,
they are necessary for the volume-integrated magnetic flux to be conserved.
The initial field is a layer of uniform flux in the $y$ direction entirely confined
below the convection zone.  The amplitude of the field is chosen so that the
volume averaged field strength is $0.5B_0$, where $B_0$ is a parameter that measures
the typical strength of the field in the radiation zone.

\begin{figure}[h!]
  \centering%
  \includegraphics{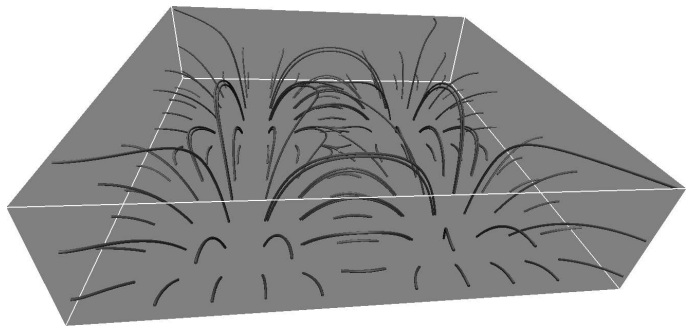}%
  \caption{%
    The initial, current-free state of the polar field simulations.
    To allow for horizontally periodic boundary conditions, the domain includes four magnetic poles.}
  \label{fig:polar-initial}%
\end{figure}

For the polar field simulations, we use the same magnetic boundary conditions
at the upper boundary,
whereas at the lower boundary we use a generalization of the conditions
used by \citet{WoodMcIntyre11}.  These boundary conditions treat the region
below the computational domain as a large reservoir of poloidal magnetic flux,
and can be concisely formulated after decomposing the magnetic field into poloidal
and toroidal scalars,
\begin{equation}
  \bB = \nablab\times\nablab\times(B_{\rm P}\mathbf{e}_z)
    + \nablab\times(B_{\rm T}\mathbf{e}_z).
  \label{eq:PTz}
\end{equation}
At the lower boundary, we impose that $B_{\rm T}$ vanishes, and we prescribe
the vertical derivative of $B_{\rm P}$.
In particular, we impose
\begin{equation}
  \dfrac{\partial B_{\rm P}}{\partial z} = \frac{4L}{\pi}B_0\sin\bigl(\frac{\pi x}{4L}\bigr)\sin\bigl(\frac{\pi y}{4L}\bigr)
  \qquad \mbox{and} \qquad
  B_{\rm T} = 0
  \qquad \mbox{at $z=2L$}.
  \label{eq:BC}
\end{equation}
These boundary conditions ensure that there is no magnetic torque at the bottom of the computational domain,
and they maintain a poloidal field of order $B_0$ through the upward diffusion of $B_{\rm P}$.
They also imply that
\begin{equation}
  B_x = B_0\cos\bigl(\frac{\pi x}{4L}\bigr)\sin\bigl(\frac{\pi y}{4L}\bigr)
  \qquad \mbox{and} \qquad
  B_y = B_0\sin\bigl(\frac{\pi x}{4L}\bigr)\cos\bigl(\frac{\pi y}{4L}\bigr)
  \qquad \mbox{at $z=2L$},
\end{equation}
so the horizontal field components are fixed, but the vertical component is unconstrained.
Because the boundaries are stress-free,
the field lines move completely freely on the boundaries.
In the absence of any flow, the field would relax to a current-free (and therefore force-free) state with
\begin{equation}
  B_{\rm P} = \frac{(4L/\pi)^2}{\sqrt{2}\cosh(\pi/\sqrt{2})}B_0\sin\bigl(\frac{\pi x}{4L}\bigr)\sin\bigl(\frac{\pi y}{4L}\bigr)\sinh\bigl(\frac{\pi z}{\sqrt{8}L}\bigr)
  \qquad \mbox{and} \qquad
  B_{\rm T} = 0.
  \label{eq:init_B}
\end{equation}
This steady state, which is plotted in \fig\ref{fig:polar-initial},
is used as the initial condition for the polar field simulations.

\subsection{The Compressible MHD Equations}

As in \citetalias{WoodBrummell12}, the ideal gas equations are
nondimensionalized using the thickness of the convective layer, $L$, as the lengthscale,
and $L/c$ as the timescale, where $c = \sqrt{p_0/\rho_0}$ is the isothermal sound speed
at the top of the domain.  The temperature, $T$, pressure, $p$, and density, $\rho$,
are nondimensionalized using $T_0$, $p_0$, and $\rho_0$, respectively.
The magnetic field, $\bB$, is nondimensionalized using $B_0$,
and diffusivities are measured in units of $Lc$.
The dimensionless ideal gas MHD equations then take the form
\begin{align}
  \rho\left(\frac{\partial}{\partial t} + \bu\cdot\nablab\right)\bu
  - 2\Omega\rho\,\mathbf{e}_z\times\bu &=
  - \nablab p
  + \frac{2}{\beta_0}(\nablab\times\bB)\times\bB
  + g\rho\mathbf{e}_z
  + 2\mu\nablab\cdot\mathbf{D}
  + \mathbf{F}
  \label{eq:mom} \\
  \left(\frac{\partial}{\partial t} + \bu\cdot\nablab\right)\rho &= -\rho\nablab\cdot\bu
  \label{eq:mass} \\
  p &= \rho T \\
  \frac{\partial\bB}{\partial t} &= \nablab\times(\bu\times\bB) + \eta\nabla^2\bB
  \label{eq:induction} \\
  \rho T\left(\frac{\partial}{\partial t} + \bu\cdot\nablab\right)\ln(p^{1/\gamma}/\rho)
  &= \nablab\cdot\left(k(z)\nablab T\right)
  + \frac{\gamma\!-\!1}{\gamma}2\mu\|\mathbf{D}\|^2
  + \frac{\gamma\!-\!1}{\gamma}\frac{2\eta}{\beta_0}|\nablab\times\bB|^2,
\end{align}
where $\gamma=5/3$ is the ratio of specific heats,
$\beta_0$ is the ratio of gas pressure to magnetic pressure, \ie
\begin{equation}
  \beta_0 = \frac{8\pi p_0}{B_0^2},
\end{equation}
$\eta$ is the dimensionless magnetic diffusivity,
and $\mathbf{D}$ is the deviatoric rate-of-strain tensor,
\begin{equation}
  D_{ij} = \frac{1}{2}\frac{\partial u_i}{\partial x_j} + \frac{1}{2}\frac{\partial u_j}{\partial x_i} - \frac{1}{3}\nablab\cdot\bu\,\delta_{ij}.
\end{equation}
The other symbols have the same meaning as in \citetalias{WoodBrummell12};
in particular,
$\Omega$ and $g$ are the constant, dimensionless
rotation rate and gravitational acceleration,
and $\mu$ and $k$ are the
dimensionless dynamic viscosity and thermal conductivity.
We take $\eta$ and $\mu$ to be constant throughout the domain, whereas for $k$
we impose a vertical profile
of the form
\begin{equation}
  k(z) \; = \; \frac{k_1}{1 + \exp(20(z-1))} \; + \; \frac{k_2}{1 + \exp(20(1-z))},
\end{equation}
so that $k = k_1$ in the upper layer, $z<1$,
and $k = k_2 > k_1$ in the lower layer, $z>1$,
the change occurring across a region
of dimensionless thickness $\simeq 0.1$.
The bottom of the convection zone is therefore fixed at $z=1$,
but convective motions are able to overshoot into the radiation zone.

We write the three components of the velocity field as $\bu = (u_x,u_y,u_z)$.
In the horizontal field simulations,
the $x$, $y$, and $z$ directions correspond to azimuth, colatitude, and depth, respectively,
and so differential rotation
is quantified by the
$x$-averaged
flow in
the $x$ direction.
In the polar field simulations,
on the other hand,
we measure the
differential rotation in terms of the large-scale vertical vorticity,
\begin{equation}
  \omega_z = \frac{\partial u_y}{\partial x} - \frac{\partial u_x}{\partial y}.
\end{equation}
Because the computational domain is horizontally symmetric
in all of our local Cartesian simulations, the Reynolds stresses
in the convective layer are not able to drive any mean
differential rotation.
In order to
mimic the generation of differential rotation in the solar convection zone, we add a
volume forcing term to the momentum \eq(\ref{eq:mom}).
In the horizontal field simulations, we use the same forcing as \citetalias{WoodBrummell12},
\begin{equation}
  \mathbf{F} = \lambda(z)\rho(\uT(y,z) - u_x)\mathbf{e}_x,
\end{equation}
where $\uT$ is the ``target'' flow
\begin{equation}
  \uT \; = \; \frac{2\Omega}{\pi}(1-z)\sin(\pi y)
\end{equation}
and $\lambda$ is the forcing rate
\begin{equation}
  \lambda \; = \; \frac{\lambda_0}{1 + \exp(20(z-1))}.
\end{equation}
In the polar field simulations, the forcing is
\begin{equation}
  \mathbf{F} = \lambda(z)\rho\big[(\uT(y,z) - u_x)\mathbf{e}_x
    + (\vT(x,z) - u_y)\mathbf{e}_y\big],
\end{equation}
where
\begin{align}
  \uT \; &= \; \frac{2\Omega}{\pi}(1-z)\sin(\pi y/2) \\
  \vT \; &= \; -\frac{2\Omega}{\pi}(1-z)\sin(\pi x/2).
\end{align}
We emphasize that $\lambda$ is exponentially small within the radiative layer $z>1$,
so that region is unforced.
In most of the simulations we take $\lambda_0 = 2\Omega$,
so that the fluid within the convective layer is pushed toward the target flow on a timescale
that is comparable to the rotation period.
In all simulations, the target flow has maximum vertical vorticity of $2\Omega$, equal to the
background vorticity of the rotating frame, implying a Rossby number of order unity in the convection zone.
We find that the mean flows established within the radiation zone
are significantly weaker than those in the convection zone, and
have a Rossby number $\ll 1$.

\subsection{Choice of Parameters}
\label{sec:parameters}
With the boundary and initial conditions listed in Section~\ref{sec:BCs},
each simulation is uniquely specified by the values of the nine dimensionless parameters
$k_1$, $k_2$, $\mu$, $g$, $\Omega$, $H$, $\eta$, $\beta_0$, and $\lambda_0$.
Each simulation presented here, unless otherwise specified, has the same non-magnetic parameters
as Case~1 of \citetalias{WoodBrummell12}, that is,
$k_1=1.45\times10^{-3}$, $k_2=2.41\times10^{-3}$, $\mu=1.45\times10^{-5}$, $g=0.24$, $\Omega=9.6\times10^{-3}$, $H=1.4\times10^{-4}$, and $\lambda_0 = 2\Omega$.
The horizontal field cases presented here therefore differ from Case~1 of \citetalias{WoodBrummell12}
only through the presence of a magnetic field,
and the fact that the domain height here is reduced to $2L$.
In all the simulations, the magnetic diffusivity is fixed at $\eta=2.9\times10^{-5}$,
and so the magnetic Prandtl number is $\mu/\eta = 0.5$.
In the next section we present three horizontal field simulations (H) and three polar field simulations (P).
The values of the two remaining parameters, $\beta_0$ and $\lambda_0$,
in each simulation
are shown in Table~\ref{tab:parameters}.
\begin{deluxetable}{ccc}

\tablecaption{Parameters for the horizontal (H) and polar (P) simulations}

\tablehead{\colhead{Case} & \colhead{$\beta_0$} & \colhead{$\lambda_0$}}

\startdata
H1 & $1.9\times10^5$ & $2\Omega$ \\
H1a & $1.9\times10^5$ & $2\Omega/5$ \\
H2 & $7.6\times10^5$ & $2\Omega$ \\
P1 & $4.8\times10^4$ & $2\Omega$ \\
P2 & $4.8\times10^5$ & $2\Omega$ \\
P3 & $4.8\times10^6$ & $2\Omega$
\enddata

\label{tab:parameters}
\end{deluxetable}

\section{Results}
\label{sec:results}

Each simulation is continued until it reaches a statistically steady state,
which in practice takes about one domain-scale magnetic diffusion time,
equivalent to about 100 rotation periods.
The results presented below are based on time averages taken over several
rotation periods in this statistically steady state.

\subsection{Horizontal Field Cases}
\label{sec:horizontal}

\subsubsection{The Formation of a Tachocline}
\begin{figure}[h!]
  \centering%
  \includegraphics{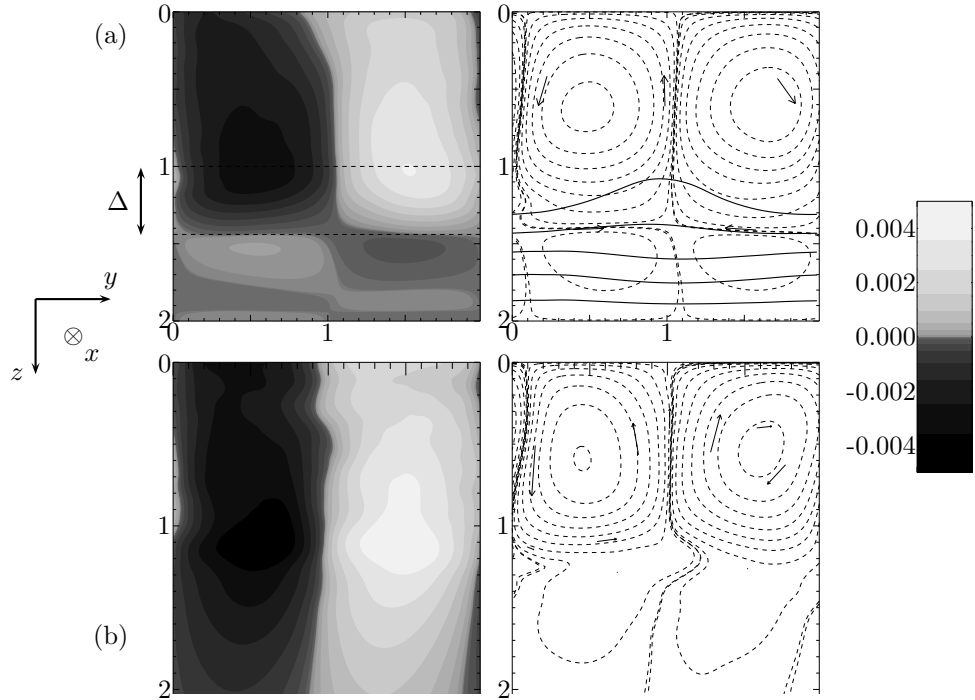}%
  \caption{%
    (a)
    Meridional cross-sections through Case H1, time-averaged over 10 rotation periods.
    (b)
    Comparable cross-sections through Case 1 of \citetalias{WoodBrummell12},
    which has no magnetic field.
    Because this latter simulation used a deeper domain, we have truncated the plots at $z=2$.
    The left panels show contours of the azimuthal flow $u_x$, using cubically spaced contour levels
    to show more detail in the radiation zone, where the flows are weakest.
    The thickness of the tachocline, $\Delta$, is indicated in the left panel of (a).
    The right panels show streamlines of the meridional flow (dashed lines and arrows) and poloidal magnetic field (solid lines).%
    }
  \label{fig:H1}%
\end{figure}

The top row of \fig\ref{fig:H1} shows the time and azimuthally averaged flow and magnetic field from Case H1.
We see that the differential rotation (\ie\ the averaged $u_x$) in the convection zone,
shown in the left panel, extends only partway into the radiation zone,
resulting in a thin tachocline.  This is in contrast
to Case~1 of \citetalias{WoodBrummell12},
shown in the bottom rom of \fig\ref{fig:H1},
in which the differential rotation was eventually
communicated all the way through the radiation zone by the meridional flow.
In Case~H1, the meridional flow that is established in the convection zone extends only to the bottom of
the tachocline, at about $z=1.44$, and a weaker counter-rotating cell develops beneath.
We choose to define the bottom of the tachocline as the depth
below the convection zone
at which the differential rotation $u_x$ first becomes zero.
The top of the tachocline is at $z=1$, where the stratification changes from adiabatic to sub-adiabatic,
and so in Case~H1 the thickness of the tachocline is approximately $\Delta \simeq 0.44$,
as indicated by dashed horizontal lines in \fig\ref{fig:H1}(a).

\begin{figure}[h!]
  \centering%
  \includegraphics{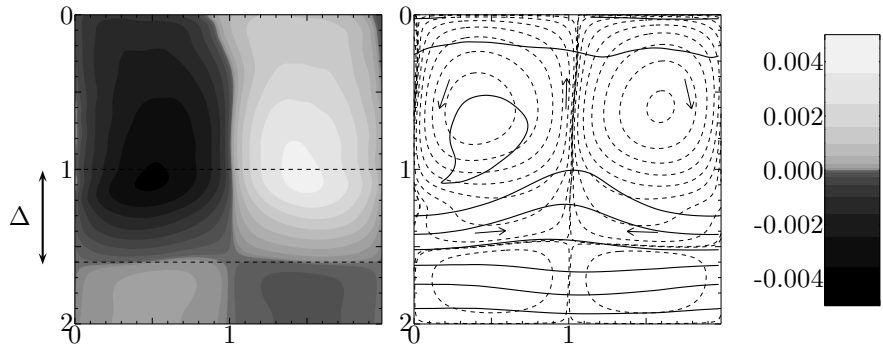}%
  \caption{%
    Same plots as \fig\ref{fig:H1}(a), but for Case~H2, and averaged over 7.5 rotation periods.
    The weaker magnetic field leads to a thicker tachocline and a deeper meridional circulation.}
  \label{fig:H2}%
\end{figure}

Since Case~H1 differs from Case~1 of \citetalias{WoodBrummell12} only through the inclusion
of a magnetic field, we must conclude that the magnetic field is responsible for the formation of
the thin tachocline in this simulation,
and that the field's strength determines the tachocline thickness.
This hypothesis is tested by Case~H2,
which has a weaker magnetic field than Case~H1,
by a factor of 2,
but is otherwise identical.
As illustrated in
\fig\ref{fig:H2},
Case~H2 has a thicker tachocline, with $\Delta \simeq 0.6$, and its meridional flows extend correspondingly deeper
into the radiation zone.
We compare these results with the predictions of several analytical models in Section~\ref{sec:thickness}.

\subsubsection{Magnetic Confinement}
\begin{figure}[h!]
  \centering%
  \includegraphics{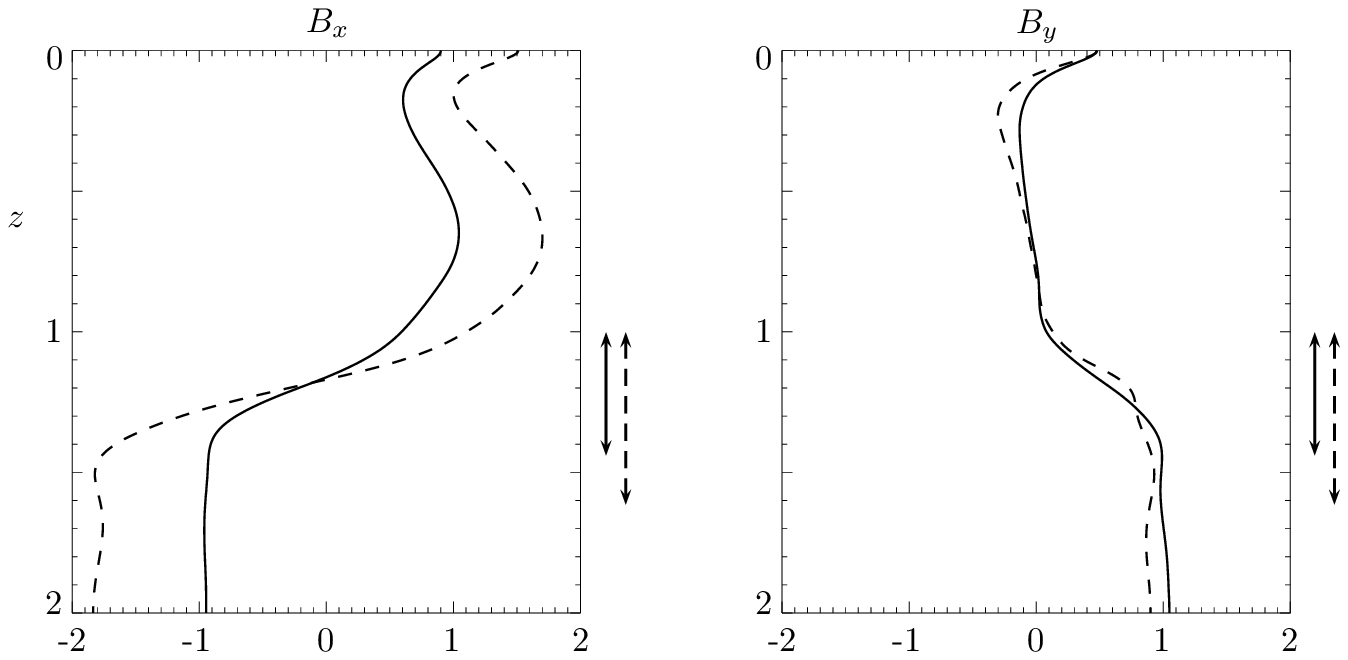}%
  \caption{%
    Time and horizontal averages of $B_x$ and $B_y$ from simulations H1 (solid lines) and H2 (dashed lines).
    Case H2 has a larger ratio of toroidal to poloidal field, but note that the magnetic scale $B_0$ is
    smaller in Case H2 by a factor of 2.
    The vertical arrows indicate the thickness of the tachocline in these simulations.}
  \label{fig:confine_H}%
\end{figure}

In both Cases H1 and H2 we find that the mean poloidal magnetic field remains confined below the
convection zone,
as can be seen in the right-hand panels of \figs\ref{fig:H1}(a) and \ref{fig:H2}.
The degree of confinement is illustrated more clearly in \fig\ref{fig:confine_H},
which shows vertical plots of the time and horizontal averages of $B_x$ and $B_y$ from
Cases~H1 and H2.
In both cases, the mean poloidal field, $B_y$, is close to zero within the bulk of the convection zone, and increases
rapidly at $z \simeq 1$ to a roughly uniform amplitude within the radiation zone.

To quantify the processes that act to confine the poloidal magnetic field, it is convenient to
adopt a poloidal--toroidal decomposition of the form
\begin{equation}
  \bB = \nablab\times(B_{\rm p}\mathbf{e}_x) + \nablab\times\nablab\times(B_{\rm t}\mathbf{e}_x).
  \label{eq:PTx}
\end{equation}
Note that this is a slightly different decomposition from that given by equation~(\ref{eq:PTz}),
but one that is more appropriate for the mean field, which is axisymmetric (\ie\ $x$-invariant).
From the induction equation (\ref{eq:induction}) we can derive an evolution equation for the
azimuthal average of the poloidal scalar $B_{\rm p}$:
\begin{align}
  \frac{\partial\overline{B}_{\rm p}}{\partial t} &= [\overline{\bu\times\bB}]_x + \eta\nabla^2\overline{B}_{\rm p} \\
  &= [\overline{\bu'\times\bB'}]_x - \overline{\bu}\cdot\nablab\overline{B}_{\rm p}
    + \eta\left(\frac{\partial^2}{\partial y^2}+\frac{\partial^2}{\partial z^2}\right)\overline{B}_{\rm p},
  \label{eq:confine}
\end{align}
where an overbar denotes an $x$ average, and primes denote departures from the average.
If the average is also taken over time in the statistically steady state, then the left-hand side of
equation~(\ref{eq:confine}) vanishes, and the three terms on the right-hand side must balance.
We identify these three terms as the transport of the mean poloidal field by waves and turbulence,
mean meridional flow, and ohmic diffusion, respectively.
In order for the field to remain confined, the first two terms together must balance the upward diffusion
of $\overline{B}_{\rm p}$.
The contribution from the mean meridional flow can be quantified by solving the induction equation (\ref{eq:induction})
kinematically with the time-averaged flow shown in \figs\ref{fig:H1}(a) and \ref{fig:H2}.
That is, we numerically integrate \eq(\ref{eq:induction}) using the time-averaged flow until the field achieves a steady state.
In both Cases H1 and H2, we find that
the steady-state poloidal magnetic field obtained is virtually identical to that shown in \figs\ref{fig:H1}(a) and \ref{fig:H2},
demonstrating that the meridional flow is primarily responsible for the field confinement in these simulations.
The confinement is produced by the ``flux expulsion'' mechanism originally described by \citet{Weiss66}:
the meridional circulation stretches out the poloidal field lines, bringing field of opposite sign into close proximity
and thereby enhancing the diffusion of the field within the convection zone.

It is perhaps surprising that the meridional flow, rather than the turbulence, dominates the transport of the mean
magnetic field in these simulations.  However, it should be remembered that the level of turbulence that can be
obtained in any numerical simulation is well below the level present in the solar convection zone,
and so our simulations almost certainly underestimate the effect of turbulent flows on the mean field.
Moreover, the manner in which we force the flow in our simulations also inhibits the turbulence
in the convection zone.
We have therefore performed an additional simulation,
which we refer to as Case H1a, which is identical to Case H1 except that
the forcing parameter $\lambda_0$ is smaller by a factor of 5.
Case H1a has stronger turbulent flows than Case H1,
and also has weaker meridional flows,
because it is the forcing in the convection zone that
ultimately ``pumps'' the meridional circulation.\footnote{Because Case H1a is more turbulent than Case H1, the numerical resolution was increased for this simulation.}

\begin{figure}[h!]
  \centering%
  \includegraphics{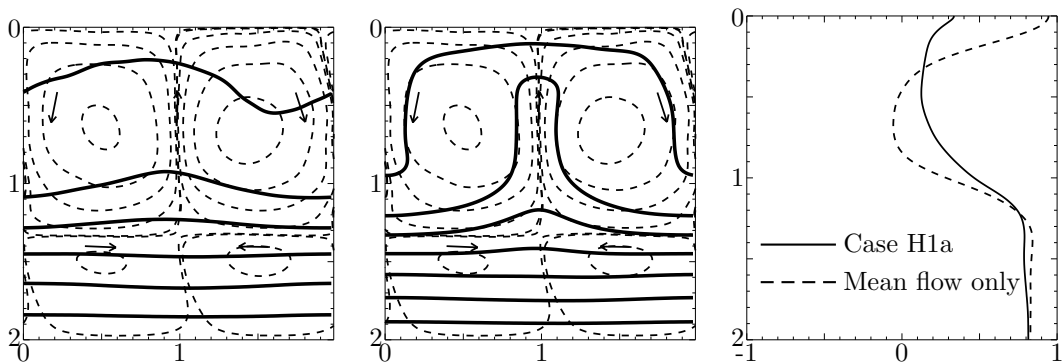}%
  \caption{%
    The mean poloidal field from Case H1a, and the steady-state field obtained by solving the induction
    equation with the mean flow only.
    The rightmost panel shows vertical profiles of the horizontal average of $B_y$.}
  \label{fig:B_expel}%
\end{figure}

In Case H1a, the mean (time and azimuthally averaged) flow contains only about 5\%
of the total kinetic energy in the simulation.
Nevertheless, in Case H1a also, we find that the geometry of the mean field is determined primarily
by the mean meridional flow,
as illustrated in \fig\ref{fig:B_expel}.
The importance of the transport by the meridional flow arises
mainly from its persistence.  Even though the meridional flow is much weaker than the turbulent flow
at each instant, the long-term behavior of the magnetic field is controlled mainly by the time-averaged flow.

This observation also explains why the degree of magnetic confinement is so similar in cases H1 and H2,
even though the tachocline in Case H2 is significantly deeper.  The field is confined not by the weak meridional
flow within the tachocline, but by the stronger meridional flows in the convection zone.
A similar result was found in the global axisymmetric model of \citet{AA-etal13}.

Although the majority of the mean poloidal field is confined to the radiation zone, we note that
some mean field also resides in a thin layer at the top of the domain.  The existence of this layer
results from the rather artificial nature of the boundary conditions used, which impose that there
is no advection, induction, or diffusion of field through the top and bottom boundaries.
In reality, this field would be mixed in with the bulk of the convection zone and, at the same time,
field would diffuse up into the tachocline from the bulk of radiation zone, maintaining a statistically-steady state.

An interesting result visible in the left panel of \fig\ref{fig:confine_H}
is that a mean toroidal field, $\overline{B}_x$, is generated, with opposite sign in the radiation and convection zones.
The source of this toroidal field is the chirality of the mean flow, and in particular the correlation between
the latitudinal gradient of the differential rotation, $\partial\overline{u}_x/\partial y$, and the vertical component of the
meridional circulation, $\overline{u}_z$.  Within the tachocline, the net effect of this correlation is an upward
transport of positive $\overline{B}_x$ and a downward transport of negative $\overline{B}_x$.
(Note that the \emph{total} toroidal field is still conserved, because of the boundary conditions.)
This result was not anticipated by earlier tachocline models
\citep{RudigerKitchatinov97,GoughMcIntyre98,Wood-etal11},
which neglected the role of the meridional flow in redistributing the toroidal magnetic field.
However, in our simulations the generation of mean toroidal field is probably enhanced by the Cartesian $f$-plane geometry,
which exaggerates the correlation between the differential rotation and the meridional flow.
Whether a similar generation of toroidal field would occur with more realistic spherical geometry
is therefore unclear.

\subsubsection{Balance of Terms}
\begin{figure}[h!]
  \centering%
  \includegraphics{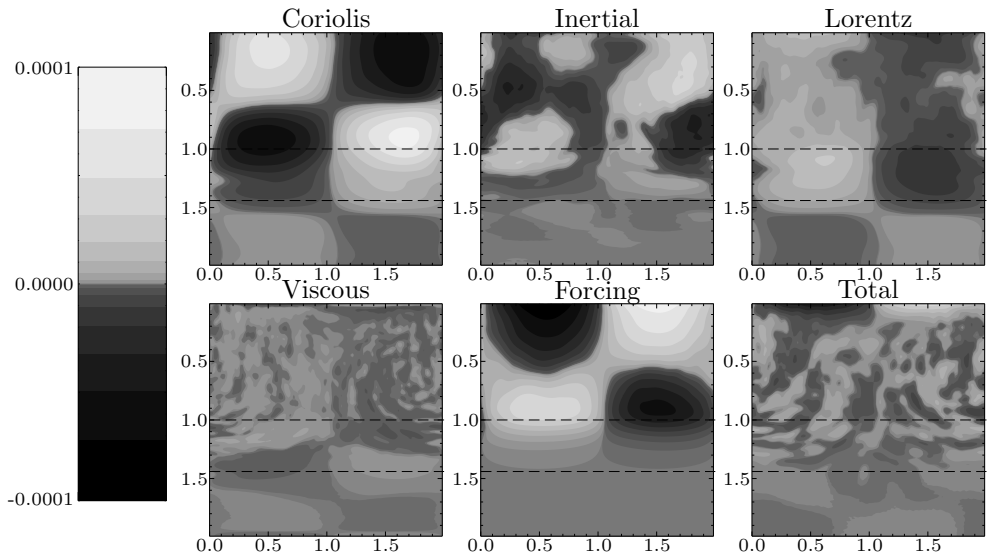}%
  \caption{%
    Each of the terms from \eq(\ref{eq:torq}), plus their total, for Case~H1.
    Contour levels are cubically spaced.}
  \label{fig:H1_torq}%
\end{figure}

The most crucial role of the magnetic field in the model of \citet{GoughMcIntyre98}
is that it balances the transport of angular momentum by the meridional flow,
which would otherwise communicate the differential rotation of the convection zone into the
radiation zone.  This balance can best be seen by taking the $x$ component of the momentum
equation (\ref{eq:mom}), after first using the continuity equation (\ref{eq:mass}) to write it in conservative form,
and taking an average over $x$ and $t$.  We then find that
\begin{equation}
  0
  =
  - 2\Omega\overline{\rho u_y}
  -\nablab\cdot(\overline{u_x\,\rho\bu})
  + \frac{2}{\beta_0}\nablab\cdot(\overline{B_x\,\bB})
  + \mu\nabla^2\overline{u}_x
  + \overline{F},
  \label{eq:torq}
\end{equation}
so in a statistically steady state there must be a balance between the mean Coriolis force, inertia, the Lorentz force,
the viscous force, and the imposed forcing.  We note that the inertial term actually contains contributions from both
mean and fluctuating fields.  However, the mean flow in the tachocline and radiation zone has a low Rossby number,
meaning that its contribution to the inertial term is negligible in comparison with the Coriolis term.
The \citeauthor{GoughMcIntyre98} model therefore predicts a balance between the Coriolis and Lorentz
terms within the tachocline.
\Fig\ref{fig:H1_torq} shows plots of each of the five terms in \eq(\ref{eq:torq}), plus their total, in Case~H1.
In the bulk of the convection zone ($0 \leqslant z \leqslant 1$),
the dominant balance is between the imposed azimuthal forcing, $F$, and the
Coriolis force from the mean latitudinal flow.  In this way, the imposed forcing ``gyroscopically pumps''
the mean meridional circulation in the convection zone, as in \citetalias{WoodBrummell12}.
However, in the radiation zone, where the forcing vanishes, the dominant balance is between the Coriolis force
and the Lorentz force.  The magnetic field thereby
enforces uniform rotation in the radiation zone, and 
prevents the ``burrowing'' of the meridional circulation seen in \citetalias{WoodBrummell12}.
We note that the viscous force is
not a dominant term anywhere in the domain.

\begin{figure}[h!]
  \centering%
  \includegraphics{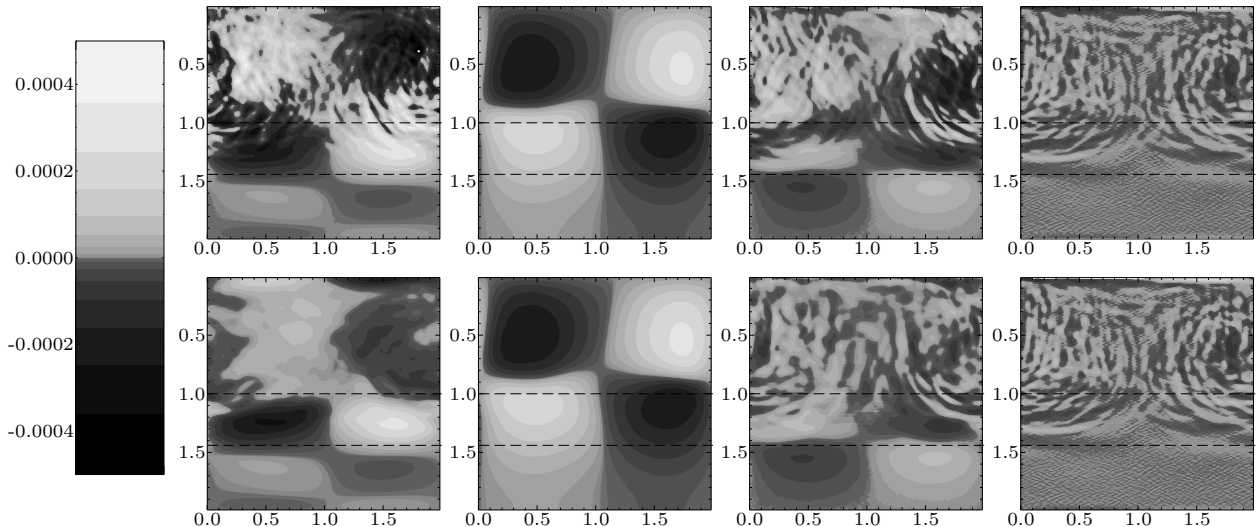}%
  \caption{%
    Top row: The contributions to the azimuthal vorticity equation in Case~H1 from the Coriolis,
    baroclinic, Lorentz, and viscous terms in \eq(\ref{eq:vorticity}).
    Bottom row: The terms in \eq(\ref{eq:vort_app}).}
  \label{fig:H1_wind}%
\end{figure}

The azimuthal momentum balance illustrated in \fig\ref{fig:H1_torq} is the same as in the
models of \citet{GoughMcIntyre98} and \citet{Wood-etal11}.
However, these two models make different predictions about the balance of forces in the
meridional directions, and in particular for the azimuthal vorticity equation.
The model of \citet{GoughMcIntyre98} assumed that this equation would closely satisfy
thermal-wind balance, \ie\ that the production of vorticity by the differential rotation would be balanced
by baroclinicity.  However, \citet{Garaud07} pointed out that if the poloidal magnetic field in the radiation zone
is sufficiently strong then the Lorentz force might modify the thermal-wind balance.
This suggestion was confirmed by \citet{Wood-etal11}, who considered stronger magnetic fields than
\cite{GoughMcIntyre98} and predicted a thinner tachopause in magneto-thermal-wind balance.
The boundary between the weak-field and strong-field regimes
can be inferred by equating the two tachopause scalings,
given by \eqs(77) and (78) in \citet{Wood-etal11}.
The result is
\begin{equation}
  \Lambda^2 = \dfrac{N^2 L^2}{2\Omega\kappa},
  \label{eq:GM_vs_WMG}
\end{equation}
where $\Lambda = |\bB|^2/(8\pi\Omega\rho\eta)$ is the Elsasser number of the magnetic field in the radiation zone,
$L$ is the horizontal lengthscale of the differential rotation, and $\kappa$ is the thermal diffusivity.
In Case~H1, the Elsasser number in the tachocline is $\Lambda \simeq 15$, and the right-hand side of
\eq(\ref{eq:GM_vs_WMG}), taking $L=1$, is approximately 200,
so we would expect a significant contribution from the Lorentz force.
To see whether this is in fact the case, we take the mean azimuthal vorticity equation in the form
\begin{equation}
  0
  =
  \nablab\cdot(\overline{u_x\boldsymbol{\omega} - \omega_x\bu})
  + \left[\overline{\nablab\times\frac{1}{\rho}\left(-\nablab p
  + \frac{2}{\beta_0}(\nablab\times\bB)\times\bB
  + 2\mu\nablab\cdot\mathbf{D}\right)}\right]_x
  \label{eq:vorticity}
\end{equation}
where $\boldsymbol{\omega} = \nablab\times\bu - 2\Omega\mathbf{e}_z$ is the absolute vorticity.
The first term in \eq(\ref{eq:vorticity})
represents the mean generation of azimuthal vorticity by vortex stretching,
and the other term incorporates the contributions from baroclinicity, Lorentz
forces, and viscous forces.  The baroclinic term is determined by the angle between the density and pressure
gradients, which are both very nearly vertical.  It is therefore common to eliminate the density in favor of entropy,
which is less dominated by its vertical gradient, at least in the convection zone \citep[\eg][]{Balbus09}.
We therefore write
\begin{align}
  \nablab\times\frac{1}{\rho}\nablab p &= \nablab\frac{1}{\rho}\times\nablab p \\
  &= \frac{1}{\rho}\nablab s\times\nablab p \\
  &\simeq g\nablab s\times\mathbf{e}_z,
\end{align}
where $s = \ln(p^{1/\gamma}/\rho)$ is the dimensionless specific entropy,
and where the last line makes the hydrostatic approximation $\nablab p \simeq \rho g\mathbf{e}_z$.
We anticipate that, below the convection zone, the mean vorticity equation (\ref{eq:vorticity})
will be dominated by contributions from the mean fields, rather than the fluctuating fields,
and can therefore be approximated as
\begin{equation}
  0 \simeq
  - 2\Omega\frac{\partial\overline{u}_x}{\partial z}
  - g\frac{\partial\overline{s}}{\partial y} +
  \nablab\cdot(\overline{\omega_\mathrm{A}}_x\overline{\mathbf{v}_\textrm{A}})
  + \frac{\mu}{\overline{\rho}}\nabla^2\overline{\omega}_x,
  \label{eq:vort_app}
\end{equation}
where
$\mathbf{v}_{\rm A} = \bB/(4\pi\rho)^{1/2}$ is the Alfv\'en velocity,
and $\boldsymbol{\omega}_{\rm A} = \nablab\times\mathbf{v}_{\rm A}$ is the Alfv\'en vorticity.
Each of the terms in \eqs(\ref{eq:vorticity}) and (\ref{eq:vort_app}) are plotted in \fig\ref{fig:H1_wind},
verifying that \eq(\ref{eq:vort_app}) does indeed closely approximate \eq(\ref{eq:vorticity})
in the tachocline and radiation zone.  We also find that, within the bulk of the tachocline,
the leading balance is between the first two terms of \eq(\ref{eq:vort_app}), \ie\ the usual
thermal-wind balance.  Below the tachocline, the Lorentz term becomes increasingly important,
whereas the viscous term is negligible everywhere, except in a thin boundary layer at the bottom
of the domain.  These results are consistent with those of \citet{Wood-etal11}: the bulk of the tachocline
is in thermal-wind balance, but the Lorentz force modifies this balance in the tachopause and beneath.
A similar result holds in Cases H1a and H2 (not shown).  In Case H2, which has a weaker magnetic field,
the contribution from the Lorentz term is somewhat weaker,
but still significant below the tachocline.

The presence of thermal-wind balance within the bulk of the tachocline allows us to predict
the distribution of entropy within that region.
Equating the first two terms in \eq(\ref{eq:vort_app}),
the vertical gradient of the differential rotation $u_x$ within the tachocline must
be balanced by a latitudinal gradient of the specific entropy $s$.
In \fig\ref{fig:H1_wind}, in particular, $s$ must be minimum at $y=1$, and maximal at $y=0,2$ within the tachocline.
Because temperature is closely correlated with entropy within the tachocline,
these entropy variations will dissipate by thermal diffusion unless they are maintained
by a combination of viscous and ohmic heating, turbulent convection, and entropy advection by mean meridional flows.
Assuming that, beneath the convection zone, the contribution from meridional flows is dominant,
we can estimate the strength of the vertical flow, $U$, required to maintain thermal-wind balance
using the method described in Section~\ref{sec:GM98}.
For the Cartesian geometry of our simulations, and using our dimensionless variables, the result is
\begin{equation}
  U \simeq \left(\frac{2\Omega k_2 L}{\rho N^2\Delta^3}\right)\left.u_x\right|_{z=1}
\end{equation}
\citep[see \eq(74) of][]{Wood-etal11},
where $N$ is the buoyancy frequency, $L$ is the horizontal scale, and $\Delta$ is the tachocline thickness.
In Case H1 we have $N\simeq0.1$, $L=1$, $u_x \simeq 0.005$ and $\Delta\simeq0.4$,
leading to a prediction of $U \simeq 4\times10^{-4}$.
At the mid-depth of the tachocline, the mean vertical velocity in Case H1 actually has an amplitude of $U \simeq 3\times10^{-4}$,
very close to the analytical prediction.

\subsubsection{The Tachocline Thickness}
\label{sec:thickness}

In summary, then, the results of the horizontal field cases are generally in agreement with the model
of \citet{Wood-etal11}, although their model was far more idealized than that used here.
A thin tachocline forms through a balance between the transport of angular momentum by
meridional circulations and magnetic fields.  The bulk of the tachocline is
in thermal-wind balance, but the Lorentz force breaks thermal-wind balance in the tachopause and beneath,
to an extent that depends on the strength of the field in the radiation zone.
The field in these simulations is confined to the radiation zone because of flux expulsion by the meridional
flows in the convection zone, even though these mean flows contain only a small fraction of
the kinetic energy.  However, the convection zone simulated here is far less turbulent, and less compressible,
that the real solar convection zone, and therefore almost certainly underestimates the contribution of
turbulent convection to the mean field transport.  Under more realistic conditions, we would expect
the turbulence to assist in the confinement of the field
\citep[\eg][]{Tobias-etal98,KitchatinovRudiger08}.

It is instructive to compare the tachocline thickness observed in Cases~H1 and H2 with that predicted by
\citet{Wood-etal11}, and with earlier tachocline models that made different assumptions about the balance of forces.
For instance, the model of \citet{RudigerKitchatinov97} neglected the role of meridional flows entirely,
instead assuming a balance between (turbulent) viscous and Maxwell stresses.
This lead to the result
\begin{equation}
  v_{\rm A} \simeq \left(\frac{\mu}{\rho\eta}\right)^{1/2}\left(\frac{L}{\Delta}\right)^2\frac{\eta}{L}\,,
\end{equation}
where $\Delta$ is the tachocline thickness,
$L$ is the horizontal lengthscale of the differential rotation,
and $v_{\rm A}$ is the Alfv\'en speed in the vicinity of the tachocline.
In our dimensionless units, with $L=1$, this corresponds to
\begin{equation}
  \Delta \simeq \left(\frac{\beta_0}{5\times10^9}\right)^{1/4}\,.
\end{equation}
The tachocline in Cases~H1 and H2 is somewhat thicker than this, reflecting the fact the
meridional flow --- and not viscosity --- dominates the transport of angular momentum in these simulations.

On the other hand, \citet{GoughMcIntyre98} predicted that
\begin{equation}
  v_{\rm A} \simeq \left(\frac{u_{\rm cz}}{\Omega L}\right)^3\left(\frac{\kappa}{\eta}\right)^{7/2}\left(\frac{\Omega}{N}\right)^7\left(\frac{L}{\Delta}\right)^9\frac{\eta}{L}\,,
\end{equation}
where
$N$ is the buoyancy frequency in the tachocline,
$\kappa = k/(\rho c_p)$ is the thermal diffusivity,
and $u_{\rm cz}$ is the differential flow velocity in the convection zone.
In our dimensionless units, with $N\simeq0.1$ and $u_{\rm cz}\simeq0.005$, this corresponds to
\begin{equation}
  \Delta \simeq \left(\frac{\beta_0}{8\times10^{11}}\right)^{1/18}\,,
\end{equation}
which is close to the thickness observed,
but does not explain the variation in $\Delta$ between Cases~H1 and H2.

Finally, \citet{Wood-etal11} predict that
\begin{equation}
  v_{\rm A} \simeq \left(\frac{u_{\rm cz}}{\Omega L}\right)\left(\frac{\kappa}{\eta}\right)\left(\frac{\Omega}{N}\right)^2\left(\frac{\Omega L^2}{\eta}\right)^{1/2}\left(\frac{L}{\Delta}\right)^3\frac{\eta}{L}
  \label{eq:WMG11}
\end{equation}
which in our dimensionless units corresponds to
\begin{equation}
  \Delta \simeq \left(\frac{\beta_0}{5\times10^7}\right)^{1/6}\,.
\end{equation}
This is roughly in accordance with the values $\Delta \simeq 0.44$ and $\Delta \simeq 0.6$ obtained in Cases~H1 and H2,
although we caution that \eq(\ref{eq:WMG11}) was obtained using a much more idealised model than that used here.
To determine whether \eq(\ref{eq:WMG11}) can reliably predict the tachocline thickness would require a more extensive parameter
study than that attempted here, and is beyond the scope of the present paper.

\subsection{Polar Field Cases}
\label{sec:polar}

The results of Section~\ref{sec:horizontal} demonstrate that the dynamics anticipated by the \citeauthor{GoughMcIntyre98} model
can be achieved in a self-consistent model, at least at those latitudes where the magnetic field geometry
is approximately horizontal.
However, closer to the (magnetic) poles
the field becomes increasingly vertical,
and can no longer be confined by flux expulsion from the convection zone.
We therefore expect the dynamics within the polar tachocline to depart significantly from those
described in the previous section.

\begin{figure}[h!]
  \centering%
  \includegraphics{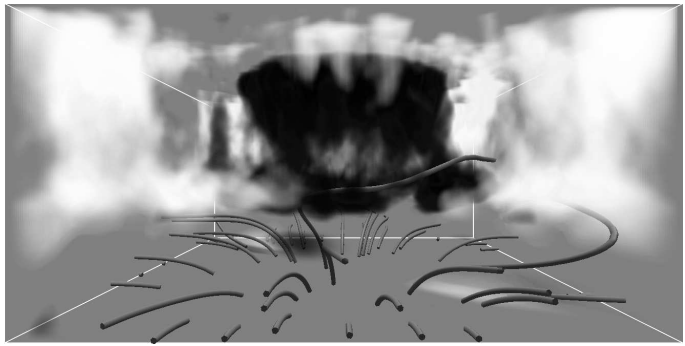}%
  \caption{%
    The time-averaged vertical vorticity, $\omega_z$, in Case~P1, which is a proxy for the differential rotation.
    Regions of significant prograde (cyclonic) and retrograde (anti-cyclonic) rotation
    are colored white and black, respectively,
    whereas regions of nearly uniform rotation are left transparent.
    A selection of field lines are also shown, whose footpoints are randomly chosen locations at the bottom of the domain.}
  \label{fig:P1}%
\end{figure}

\Fig\ref{fig:P1} shows the the time-averaged differential rotation and magnetic field in Case~P1,
averaged over 7 rotation periods.
We find that the differential rotation
is mainly confined to the convection zone,
with a concentration of retrograde rotation (black) over the pole
surrounded by prograde rotation (white).
By contrast, the radiation zone rotates uniformly,
and a thin tachocline is established at their interface, $z=1$.
Compared to the initial equilibrium field (\fig\ref{fig:polar-initial}),
the magnetic field lines are much more horizontal,
and are mostly confined to the radiation zone.

\begin{figure}[h!]
  \centering%
  \includegraphics{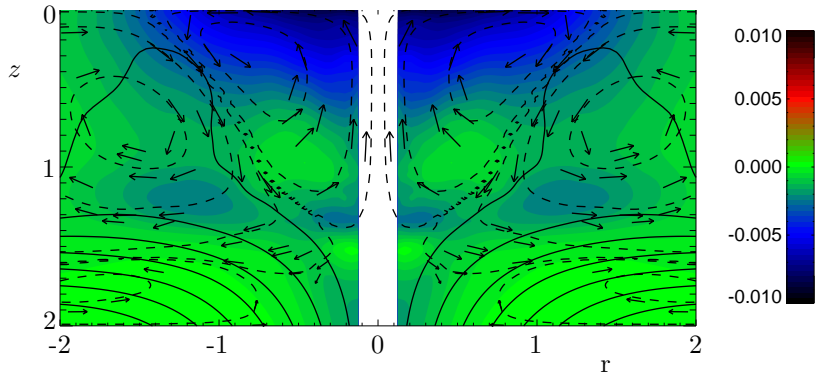}%
  \caption{%
    Meridional cross-section showing
    the mean poloidal magnetic field (solid lines), meridional circulation (dashed lines and arrows),
    and angular velocity (color scale) in Case~P1, after averaging in azimuth.
    The angular velocity on the rotation axis is ill-defined, and therefore not shown.%
    }
  \label{fig:P1_axi}%
\end{figure}

The artificial Cartesian geometry of the polar simulations
makes it difficult to meaningfully define the azimuthal average of the fields,
because the mean fields retain significant non-axisymmetric features even in a long-time average.
Nevertheless, it is still useful to extract the axisymmetric component of the mean fields;
this can be achieved efficiently by taking the Fourier transform in the $x$ and $y$ directions and then
calculating the first term in the Jacobi--Anger expansion for each Fourier mode.
The result for Case~P1 is shown in \fig\ref{fig:P1_axi} in a neighborhood of the pole.
The confinement of the magnetic field is evident in this plot;
with the exception of a few field lines very close to the rotation axis,
the mean field is entirely confined below the convection zone.
The radiation zone is in almost uniform rotation,
apart from some slightly retrograde rotation very close to the rotation axis,
where the field lines are not confined.
However, it should be noted that the mean rotation rate becomes rather ill-defined close to the rotation axis,
particularly in the convection zone, where time-dependent velocity fluctuations greatly exceed the mean.
For this reason, it is much harder to measure the thickness of the tachocline in the polar simulations
than in the horizontal field simulations presented earlier.
\citep[Of course, the same difficulty arises in measuring the thickness of the solar tachocline close to the poles, \eg][]{BasuAntia03}.
Based on a purely visual inspection of \fig\ref{fig:P1_axi},
we estimate the bottom of the tachocline to be at a depth of around $z \simeq 1.4$.
Finally, we note that the meridional flow is downwelling in the tachocline throughout the region
shown in \fig\ref{fig:P1_axi}, as expected.  The meridional flow in the convection zone, on the other hand,
is more complex and is actually upwelling over the pole.

\begin{figure}[h!]
  \centering%
  \includegraphics{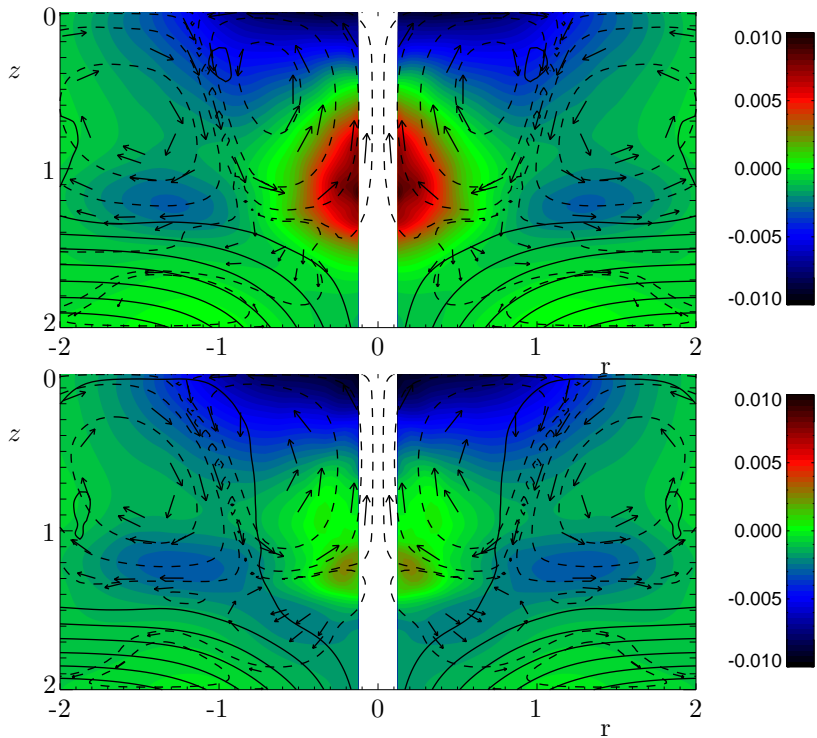}%
  \caption{%
    Meridional cross-sections comparable to \fig\ref{fig:P1_axi},
    but for Cases P2 and P3.%
    }
  \label{fig:P23_axi}%
\end{figure}
\begin{figure}[h!]
  \centering%
  \includegraphics{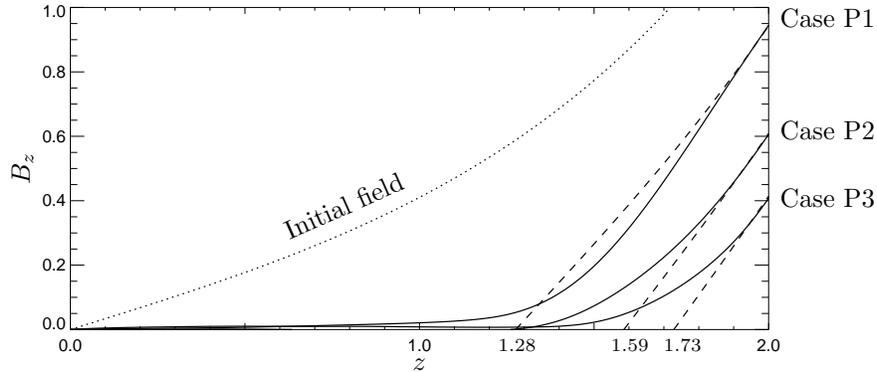}%
  \caption{%
    Solid lines: amplitude of the Fourier mode of $B_z$ with $k_x = k_y = \tfrac{\pi}{4}$
    in Cases P1, P2, and P3.
    Dashed lines: the vacuum solution (\ref{eq:vacuum}),
    with $z_0$ chosen to match the true solution at $z=2$.
    Dotted line: the initial, unconfined vacuum solution (\ref{eq:init_B}).%
    }
  \label{fig:vacuum}%
\end{figure}

Because of the difficulty in defining the azimuthal average in our cartesian polar simulations,
it is difficult to quantitatively analyze the results in the same way as in the preceding section.
But by comparing results from simulations performed at different parameters
we can still determine how the thickness of the tachocline,
and the degree of magnetic confinement,
depend on the strength of the primordial field.
We therefore present, in \fig\ref{fig:P23_axi}, Cases P2 and P3, which have weaker magnetic fields than Case P1,
but are otherwise identical.
These plots highlight the issue mentioned above, that the rotation rate of the convection zone close to the axis
fluctuates enormously, and the mean rotation rate is not clearly defined even after averaging over several rotation
periods.
The most conspicuous differences in the angular velocity in Figures~\ref{fig:P1_axi} and \ref{fig:P23_axi}
result from differing patterns of convective cells, rather from significant differences in the dynamics.
Nevertheless, these plots demonstrate that if the primordial magnetic field is made weaker
then the mean field lines become more deeply confined within the radiation zone.
In order to more precisely define the degree of magnetic confinement in these simulations,
we first note that, far below the convection zone, the flows become very weak and so the mean magnetic field
must converge asymptotically to a potential field with $\nabla^2B_z = 0$.
To meet the lower boundary condition (\ref{eq:BC}),
the field must therefore have the form
\begin{equation}
  B_z \simeq \sqrt{2}\sin(\tfrac{\pi}{4}x)\sin(\tfrac{\pi}{4}y)\frac{\sinh(\tfrac{\pi}{\sqrt{8}}(z-z_0))}{\cosh(\tfrac{\pi}{\sqrt{8}}(2-z_0))}
  \label{eq:vacuum}
\end{equation}
near the bottom of the domain, for some constant $z_0$.
We will refer to $z_0$ as the ``effective confinement depth'', because this is the depth at which the magnetic field
would become horizontal if the vacuum solution (\ref{eq:vacuum}) were continued up toward the convection zone.
We note that the initial condition for the polar simulations (\ref{eq:init_B})
has a confinement depth of $z_0=0$, because the boundary conditions at $z=0$ impose that the field is horizontal there.
To quantify the degree of magnetic confinement we therefore
take the Fourier component of $B_z$ with horizontal wave numbers $k_x = k_y = \tfrac{\pi}{4}$ at the bottom of
the domain, $z=2$,
and use it to compute the value of $z_0$ in \eq(\ref{eq:vacuum}).
The result is illustrated in
\fig\ref{fig:vacuum}, which shows the amplitude of this Fourier mode as a function of depth for Cases P1, P2 and P3,
and the corresponding vacuum solutions after matching at $z=2$.
The effective confinement depths for the three cases are 1.28, 1.59, and 1.73 respectively,
demonstrating that a weaker field is confined deeper within the radiation zone.
For reference, the dotted line in \fig\ref{fig:vacuum} shows the initial condition (\ref{eq:init_B}).
\begin{figure}[h!]
  \centering%
  \includegraphics{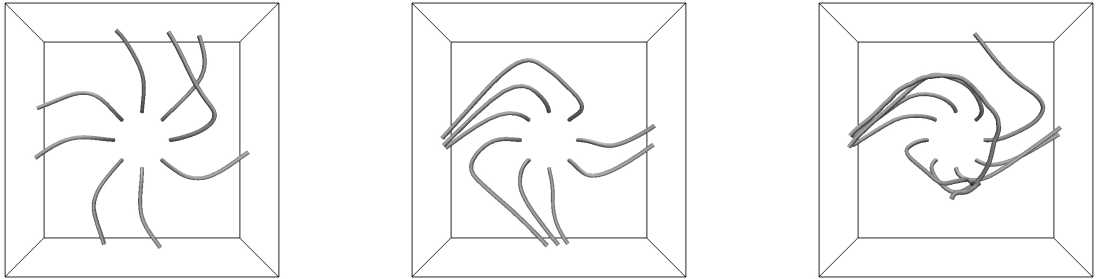}%
  \caption{%
    Plan view of the time-averaged magnetic field lines in Cases P1, P2, and P3.%
    }
  \label{fig:polar_plan}%
\end{figure}

\Fig\ref{fig:polar_plan} shows plan views of the time-averaged magnetic field lines in Cases P1, P2, and P3,
demonstrating that a weaker poloidal field is more tightly wound up by the differential rotation in the tachocline.
The same result was found by \cite{WoodMcIntyre11}
in their axisymmetric model.
The analogue of
\eq(\ref{eq:torq}) in the polar tachocline, assuming an axisymmetric balance between the mean Coriolis and Lorentz torques,
is
\begin{equation}
  -2\Omega ru_r = \frac{2}{\beta_0}\bB_{\rm p}\cdot\nablab(rB_\phi)
\end{equation}
where $\bB_{\rm p}$ is the poloidal magnetic field.
In each of the polar simulations, the strength of the meridional flow, $u_r$, is similar, and
so a reduction in the strength of $\bB_{\rm p}$ must be compensated for by an increase in $B_\phi$,
in order to maintain this balance.

The fact that weaker magnetic fields are more deeply confined near the pole --- in contrast to the situation for horizontal
magnetic fields seen in \fig\ref{fig:confine_H} --- suggests that near the pole the degree of magnetic confinement
is determined primarily by the meridional flow in the tachocline.
To confirm this hypothesis we have performed an additional simulation
in which the forcing in the convection zone,
and therefore the differential rotation and meridional flows, are switched off.
This simulation was initialized with the confined magnetic field realized in Case P1.
On the timescale of magnetic diffusion across the domain
the field became increasingly unconfined,
eventually resembling the vacuum force-free magnetic field (\ref{eq:init_B}).
We conclude that the turbulence in the convection zone cannot by itself confine the magnetic field,
and so the magnetic confinement achieved in Cases P1, P2, and P3 can be attributed to the mean downwelling flow
in the tachocline.

\section{Summary and conclusions}
\label{sec:summary}

We have presented a fully nonlinear 3D numerical model of the solar tachocline.
These are the first numerical simulations in which a tachocline forms self-consistently at the interface
between the convection and radiation zones,
and remains thin on long timescales (including the timescale of viscous diffusion across the domain).
Uniform rotation is maintained in the radiation zone by a confined primordial magnetic field,
whose
azimuthal Lorentz force balances the Coriolis force from the mean meridional flow.
In the absence of the magnetic field, the tachocline would thicken and ultimately extend deep
into the radiation zone, as in the simulations of \citetalias{WoodBrummell12}.
At the same time, the mean meridional flow confines the magnetic field below the convection zone,
as originally envisaged by \citet{GoughMcIntyre98}.
Crucially, viscous stresses are subdominant in the tachocline and beneath,
so our results demonstrate the magnetostrophic balance that is expected to hold in the real solar tachocline.

We have considered two different magnetic field geometries,
to represent the conditions within the tachocline at different latitudes.
The horizontal field cases presented in Section~\ref{sec:horizontal}
qualitatively resemble the laminar axisymmetric models of \citet{GoughMcIntyre98} and \citet{Wood-etal11},
and the balance of forces roughly agree with the results of those models.
In particular, the tachocline is approximately in thermal-wind balance,
with latitudinal variations in entropy that are maintained against thermal diffusion by the meridional flow.
In the ``tachopause'' at the bottom of the tachocline, and in the layers beneath,
thermal-wind balance is modified by the Lorentz force.
In these simulations the confinement of the magnetic field is produced by flux expulsion by the meridional flow
in the convection zone,
even though the meridional flow is much weaker than the turbulent convection in that region.

The polar field cases presented in Section~\ref{sec:polar} closely resemble
the axisymmetric polar model of \citet{WoodMcIntyre11}.
The magnetic field is confined by the downwelling meridional flow in the tachocline,
whose penetration depth into the radiation zone depends on the strength of the magnetic field.
A weaker magnetic field is thus more deeply confined, producing a thicker tachocline.

In all of our simulations, the confinement of the field can be attributed entirely to the mean meridional flow.
However, in the real Sun, convective turbulence
is also expected to play an important role in the transport of the mean field,
through flux expulsion
\citep{Zeldovich57,Weiss66,Radler68,KitchatinovRudiger92,Tao-etal98}
and topological pumping
\citep{DrobyshevskiYuferev74,Tobias-etal98,DorchNordlund01,KitchatinovRudiger08}.
The absence of significant pumping in our simulations is probably explained by the relatively small density
contrast across the domain, which is a limitation of our local model.
Ultimately, global simulations are required to determine how the tachocline dynamics
described here are modified by the transport of magnetic flux, and angular momentum,
within the convection zone.
Our simulations can inform the choice of parameters in such global models.

We thank
Pascale Garaud,
Gary Glatzmaier,
C\'eline Guervilly,
Michael McIntyre,
and the anonymous referee
for useful discussions and suggestions.
T.S.W.~was supported by NSF CAREER grant 0847477.
N.H.B.~was supported by NASA grants NNX07AL74G and NNX14AG08G,
and by the Center for Momentum Transport and Flow Organization (CMTFO),
a DoE Plasma Science Center.
Numerical simulations were performed on NSF XSEDE resources Kraken and Ranger,
supported by NSF grant ACI-1053575,
and the Pleiades supercomputer at University of California Santa Cruz purchased under
NSF MRI grant AST-0521566.

\end{document}